\documentclass[12pt,prd,showpacs,tightenlines,nofootinbib]{revtex4}
\usepackage{bm}
\usepackage{graphics}
\usepackage{rotating}
\usepackage{epsfig}
\usepackage{slashed}
\begin{document}
\title{
Rare $\Lambda_c\to p \ell^+\ell^-$  decay in
the relativistic quark model}  
\author{R. N. Faustov}
\author{V. O. Galkin}
\affiliation{Institute of Cybernetics and Informatics in Education, FRC CSC RAS,
  Vavilov Street 40, 119333 Moscow, Russia}

\begin{abstract}
The relativistic quark model based on the quasipotential approach with
the QCD-motivate potential is
employed for the calculation of the form factors of the $\Lambda_c\to
p$ rare weak transitions. Their momentum dependence is explicitly
determined without additional assumptions and extrapolations in the
whole kinematical range of the momentum transfer squared $q^2$. The
differential $\Lambda_c\to p l^+l^-$ decay branching fractions and
angular distributions are calculated on the basis of these form
factors. Both the perturbative and effective Wilson coefficients,
which include contributions of vector meson resonances, are used. The
calculated branching fraction of the $\Lambda_c\to p \mu^+\mu^-$ rare
decay is well consistent with the experimental upper limit very recently
set by the LHCb Collaboration.

\end{abstract}


\maketitle

\section{Introduction}

In the standard model the $\Lambda_c\to p \ell^+\ell^-$ rare weak decays are governed by the
$c\to u$ quark transitions which proceed through the flavour-changing neutral
currents.  The short-distance contributions to these decays are
expected to be strongly suppressed by the Glashow-Iliuopoulos-Maiani
(GIM) mechanism \cite{gim}. Indeed, the corresponding penguin diagrams get
contributions only from the down-type quarks which have negligible masses
compared to the electroweak scale thus providing almost complete GIM
cancellation.   Therefore a significant role is played by the long-distance
effects which are expected to arise from the   meson
resonances decaying to the lepton pair. As a result process involving
$c\to u l^+l^-$ transitions are far less explored than the
corresponding $b\to s l^+l^-$ transitions both theoretically and
experimentally. The exclusive decays proceeding
through such transitions are mostly studied for the rare $D$-meson
decays (see e.g. recent papers \cite{fk,bh,fms} and references therein). The baryon
case  $\Lambda_c\to p l^+l^-$ received substantially less
attention. The estimates of the corresponding form factors and decay
branching fractions in the light cone QCD sum rules are given in
Refs.~\cite{abss1,s}. The lattice QCD determination of
the $\Lambda_c\to p$ form factors and the $\Lambda_c\to p \mu^+\mu^-$ rare
decay observables was recently presented in Ref.~\cite{latt}. Experimentally
first constraints on the  $\Lambda_c$ rare decay branching fractions
were set by the BABAR Collaboration \cite{babarLc} and very recently
they were significantly improved by the LHCb Collaboration
\cite{lhcblc}.  

In this paper we calculate the  $\Lambda_c\to p$ transition form
factors in the framework of the relativistic quark model based on
quasipotential approach with QCD-motivated potential. We consider
baryons to be the relativistic quark-diquark bound systems which wave
functions were previously determined within the mass spectra calculations
\cite{barregge}. Using the quasipotential approach we express the weak
transition matrix elements through the overlap integrals of the
initial and final baryon wave functions. It is important to emphasize
that the momentum transfer dependence of the matrix elements and
corresponding form factors  is explicitly determined
without additional assumptions and extrapolations in the whole available kinimatical
range. We previously successfully applied such approach for the study
of the semileptonic and rare $\Lambda_b$ decays \cite{lbdecay,lbrare}
and the semileptonic $\Lambda_c$ decays \cite{lcdecay}. Then we use these
form factors for the calculation of the differential and total decay
branching fractions both with the perturbative and effective Wilson
coefficients, which include the long-distance contributions from
vector meson resonances, and confront the obtained results with the
available experimental data.

\section{Relativistic quark model}

We employ the relativistic quark-diquark picture based on the
quasipotential approach for the description of baryon properties. The
diquark as a bound state of two quarks and the baryon as a quark-diquark
bound system are described by the
diquark wave function $\Psi_{d}$
and by the baryon wave function $\Psi_{B}$, which satisfy the relativistic
quasipotential equation of the Schr\"odinger type \cite{mass}
\begin{equation}
\label{quas}
{\left(\frac{b^2(M)}{2\mu_{R}}-\frac{{\bf
p}^2}{2\mu_{R}}\right)\Psi_{d,B}({\bf p})} =\int\frac{d^3 q}{(2\pi)^3}
 V({\bf p,q};M)\Psi_{d,B}({\bf q}).
\end{equation}
Here the relativistic reduced mass is defined as
\[
\mu_{R}=\frac{M^4-(m^2_1-m^2_2)^2}{4M^3},\] 
 and the center-of-mass system
relative momentum squared on mass shell is given by 
\[{b^2(M) }
=\frac{[M^2-(m_1+m_2)^2][M^2-(m_1-m_2)^2]}{4M^2},
\]
where $M$ is the bound state mass (diquark or baryon),
$m_{1,2}$ are the masses of  quarks ($q_1$ and $q_2$) which form
the diquark or of the  diquark ($d$) and  quark ($q$) which form
the baryon ($B$), and ${\bf p}$  is their relative momentum.

To construct the quasipotentials $V({\bf p,q};M)$  of
the quark-quark or quark-diquark interaction we use the off-mass-shell scattering amplitude, projected onto the positive
energy states. The effective quark interaction is taken to be the
sum of the one-gluon exchange term and the mixture of long-range
vector and scalar linear confining potentials with the mixing
coefficient $\varepsilon$. In the nonrelativistic limit they are given by
\begin{eqnarray}
V_{\rm OG}(r)&=&-\frac43\frac{\alpha_s}{r}, \cr
V^V_{\rm conf}(r)&=&(1-\varepsilon)(Ar+B),\cr
V^S_{\rm conf}(r) &=&\varepsilon (Ar+B),
\end{eqnarray} 
and their sum \[V(r)=V_{\rm OG}(r)+V^V_{\rm conf}(r)+ V^S_{\rm conf}(r)=-\frac43\frac{\alpha_s}{r}+Ar+B\]
reproduce  the widely used Cornell-like potential. Note that we use the freezing \cite{ltetr} QCD coupling constant $\alpha_s$. 
As in the case of mesons \cite{mass}, we  also assume that
the vector confining potential contains not only the Dirac term but
the additional Pauli term, thus introducing the anomalous chromomagnetic quark
moment $\kappa$:
\begin{equation}
\Gamma_{\mu}({\bf k})=\gamma_{\mu}+
\frac{i\kappa}{2m}\sigma_{\mu\nu}\tilde k^{\nu}, \qquad \tilde
k=(0,{\bf k}).
\end{equation} 
The explicit expressions for the
quasipotentials  are given in Ref.~\cite{hbar}.

All parameters of the model were
fixed previously from calculations of meson and baryon properties
\cite{mass,hbar}. We use the following values for the constituent quark masses: $m_u=m_d=0.33$
GeV, $m_s=0.5$ GeV, $m_c=1.55$ GeV  and for the parameters of the linear potential:
$A=0.18$ GeV$^2$ and $B=-0.3$ GeV.  The value of the mixing coefficient of vector and scalar
confining potentials $\varepsilon=-1$ has been determined from the
consideration of the heavy quark expansion for the semileptonic heavy
meson decays and the charmonium radiative decays \cite{mass}. While the
universal Pauli interaction constant $\kappa=-1$  has been fixed from
the analysis of the fine splitting of heavy quarkonia ${}^3P_J$-
states \cite{mass}.  Note that the long-range chromomagnetic
contribution to the potential, which is proportional to $(1+\kappa)$,
vanishes for the chosen value of $\kappa=-1$ in agreement with the
flux tube model.

\section{Form factors of the rare weak $\Lambda_c\to p$ transitions}

To study the rare weak decays of the $\Lambda_c$ baryon we need to
calculate the  matrix element of the  weak current between the $\Lambda_c$ and
proton ($p$). This matrix element in the quasipotential approach is given by the expression  
\begin{equation}\label{mxet} 
\langle p(p_{p}) \vert J^W_\mu \vert \Lambda_c(p_{\Lambda_c})\rangle
=\int \frac{d^3p\, d^3q}{(2\pi )^6} \bar \Psi_{p\,{\bf p}_{{p}}}({\bf
p})\Gamma _\mu ({\bf p},{\bf q})\Psi_{\Lambda_c\,{\bf p}_{\Lambda_c}}({\bf q}),
\end{equation}
where $\Gamma _\mu ({\bf p},{\bf
q})$ is the two-particle vertex function and  
$\Psi_{B\,{\bf p}_{B}}$ are the $B$ ($B=\Lambda_c,p$) 
baryon  wave functions projected onto the positive-energy states of 
quarks. The vertex function $\Gamma$ receives
relativistic contributions both from the impulse approximation diagram
and from the
diagrams with the intermediate negative-energy states
\cite{lbdecay}. Since the final proton is moving with the momentum ${\bf
  p}_{p}$ in the $\Lambda_c$ rest frame (${\bf p}_{\Lambda_c}=0$) the boosts of the proton wave
function are taken in to account by the wave function transformation \cite{lbdecay} 
 \begin{equation}
\label{wig}
\Psi_{{p}\,{\bf p }_p}({\bf
p})=D_{u}^{1/2}(R_{L_{{\bf p}_p}}^W)\Psi_{{p}\,{\bf 0}}({\bf p}),
\end{equation}
where $\Psi_{{p}\,{\bf 0}}$ is the proton wave function in the
rest frame,  $R^W$ is the Wigner rotation, $L_{\bf p}$ is the Lorentz boost
from the baryon rest frame to a moving one with the momentum ${\bf p}_p$, and   
 $D^{1/2}_u(R^W)$ is the rotation matrix of the active ($u$) quark
 spin. Note that we consider a proton as the bound state of the $u$ quark and
 scalar $ud$ diquark.

The hadronic matrix elements for the weak decay $\Lambda_c\to
p\ell^+\ell^-$  can be parameterized by the following set of the
invariant form factors \cite{giklsh,gikls}
\begin{eqnarray}
  \label{eq:ff}
&&\!\!\!\!\!\!  \langle p(p',s')|\bar{u} \gamma^\mu c|\Lambda_c(p,s)\rangle= \bar
  u_{p}(p',s')\Bigl[f_1^V(q^2)\gamma^\mu-f_2^V(q^2)i\sigma^{\mu\nu}\frac{q_\nu}{M_{\Lambda_c}}+f_3^V(q^2)\frac{q^\mu}{M_{\Lambda_c}}\Bigl]
u_{\Lambda_c}(p,s),\cr
&&\!\!\!\!\!\!\! \langle p(p',s')|\bar{u} \gamma^\mu\gamma_5 c|\Lambda_c(p,s)\rangle= \bar
  u_{p}(p',s')[f_1^A(q^2)\gamma^\mu-f_2^A(q^2)i\sigma^{\mu\nu}\frac{q_\nu}{M_{\Lambda_c}}+f_3^A(q^2)\frac{q^\mu}{M_{\Lambda_c}}\Bigl]
\gamma_5 u_{\Lambda_c}(p,s),\qquad \cr
&&\!\!\!\!\!\!\langle p(p',s') | \bar{u} i\sigma^{\mu\nu}q_\nu c | \Lambda_c(p,s)
                                                                        \rangle = \bar{u}_p(p',s') \left[  \frac{f_1^{TV}(q^2)}{M_{\Lambda_c}} \left(\gamma^\mu q^2 - q^\mu \slashed{q} \right) - f_2^{TV}(q^2) i\sigma^{\mu\nu}q_\nu  \right] u_{\Lambda_c}(p,s), \cr 
&&\!\!\!\!\!\!\!\!\!\!\!\!\!\langle p(p',s') | \bar{u}
   i\sigma^{\mu\nu}q_\nu\gamma_5 c|\Lambda_c(p,s)\rangle
  = \bar{u}_p(p',s') \left[  \frac{f_1^{TA}(q^2)}{M_{\Lambda_c}}
      \left(\gamma^\mu q^2 - q^\mu \slashed{q} \right) - f_2^{TA}(q^2)
      i\sigma^{\mu\nu}q_\nu  \right]\gamma_5 u_{\Lambda_c}(p,s).
\end{eqnarray}

Explicit expressions for the corresponding form factors
in our model are given in Refs. \cite{lbdecay,lbrare}. The form factors are
expressed through the overlap integrals of the baryon wave
functions which we take from the mass spectrum calculations \cite{barregge}. All
relativistic effects including transformations of the proton wave
functions from the rest to moving reference frame (\ref{wig}) and contributions of
the intermediate negative-energy states are consistently taken into account. It is
important to point out that the momentum transfer  $q^2$ behavior is
explicitly determined in the whole kinematical range without
extrapolations or additional assumptions which are used in most of other
theoretical considerations. This fact improves the reliability of the form
factor calculations. 

\begin{figure}
\centering
  \includegraphics[width=8cm]{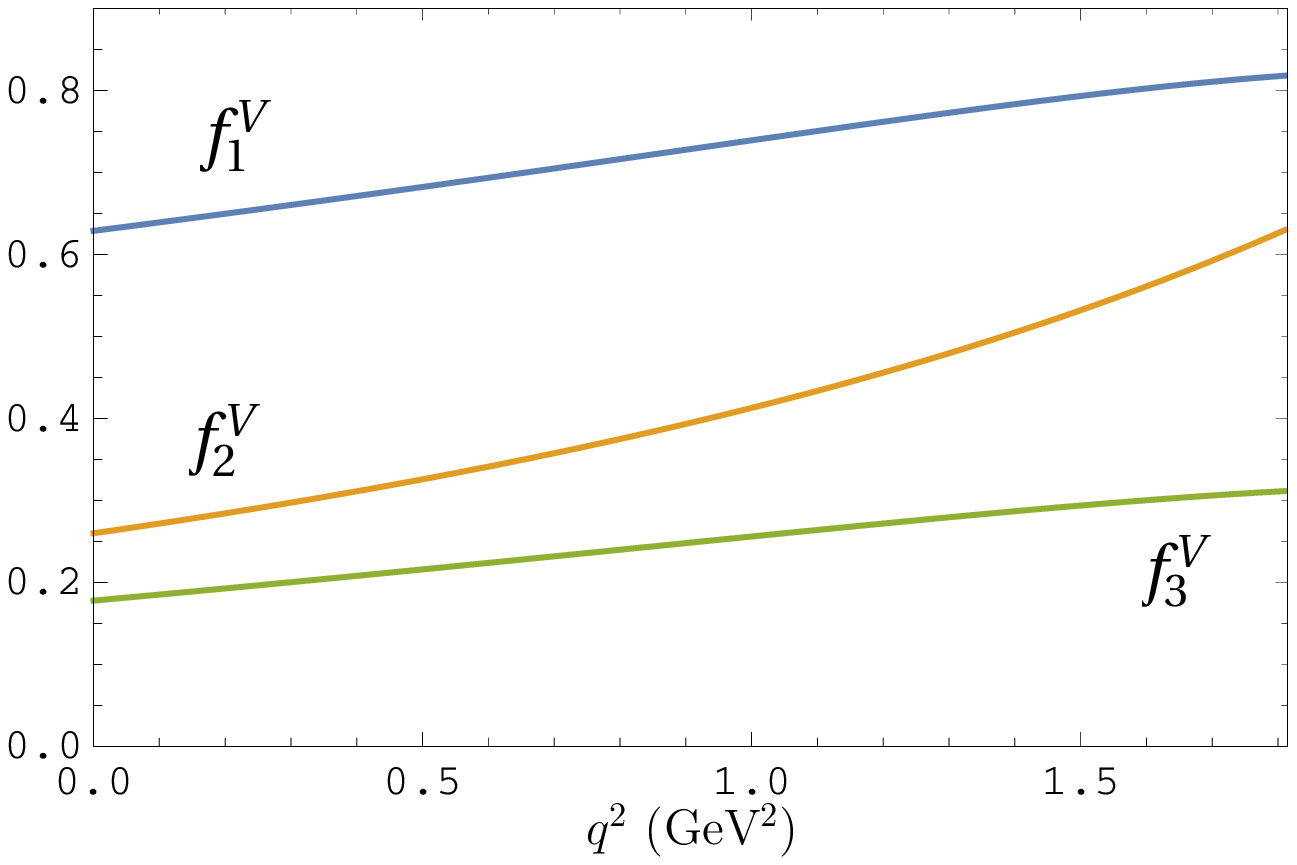}\ \
 \ \includegraphics[width=8cm]{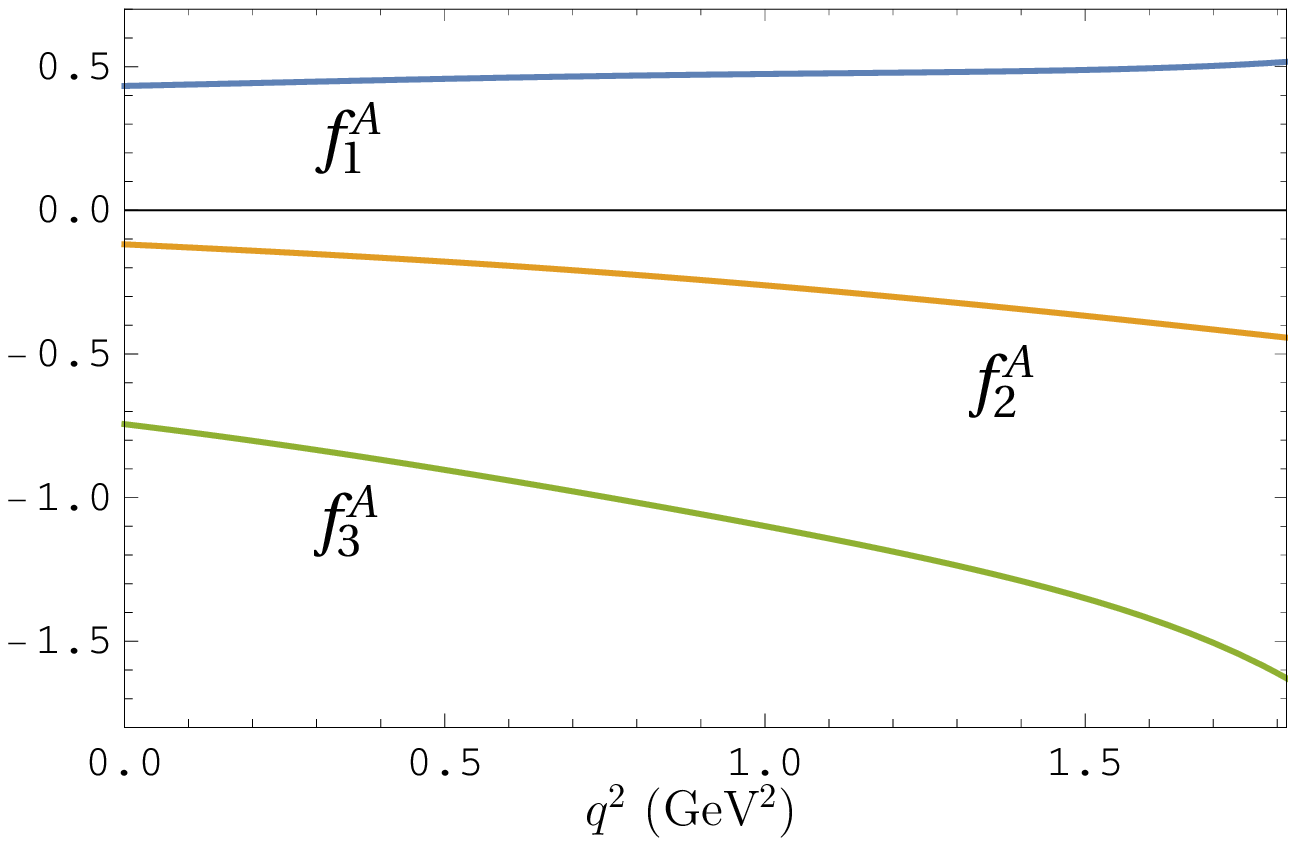}\\
\includegraphics[width=8cm]{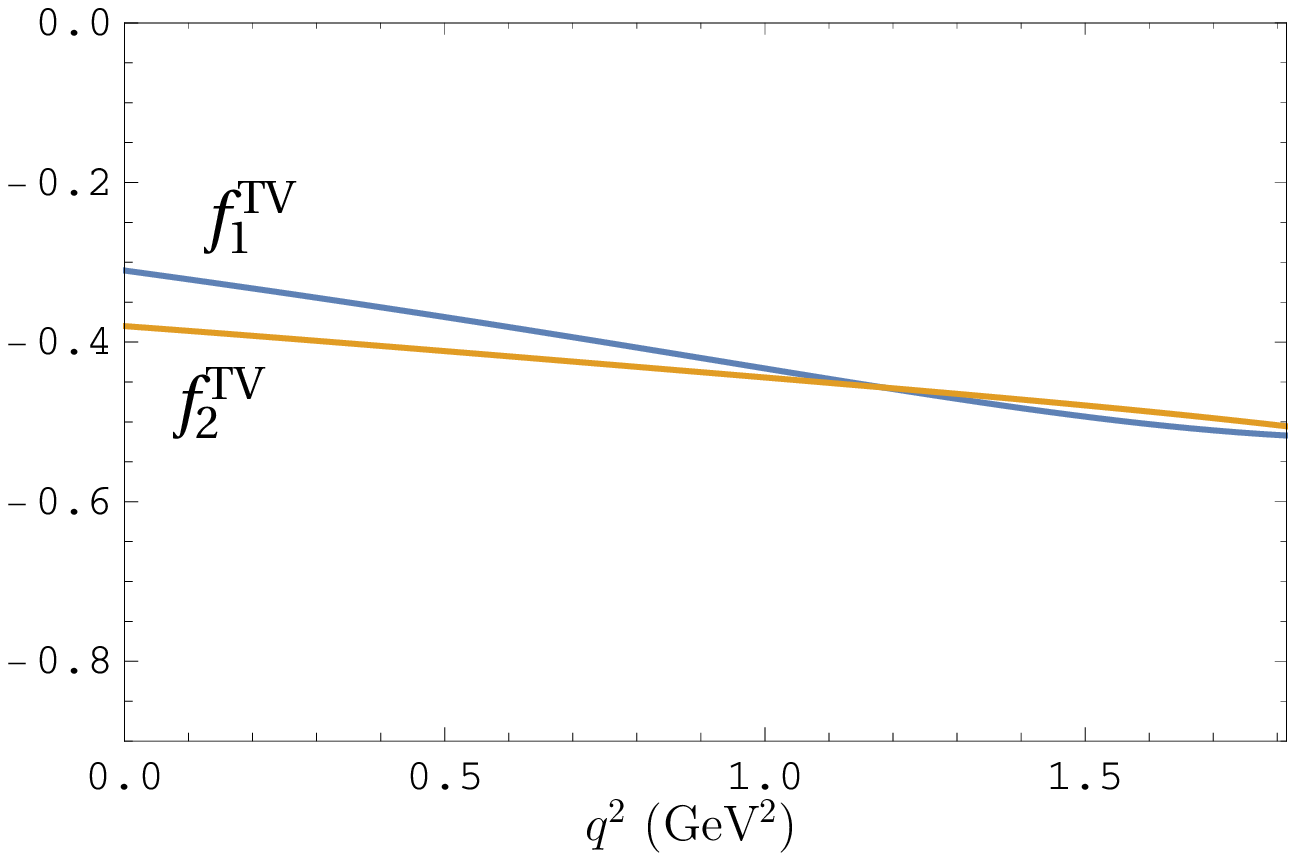}\ \
 \ \includegraphics[width=8cm]{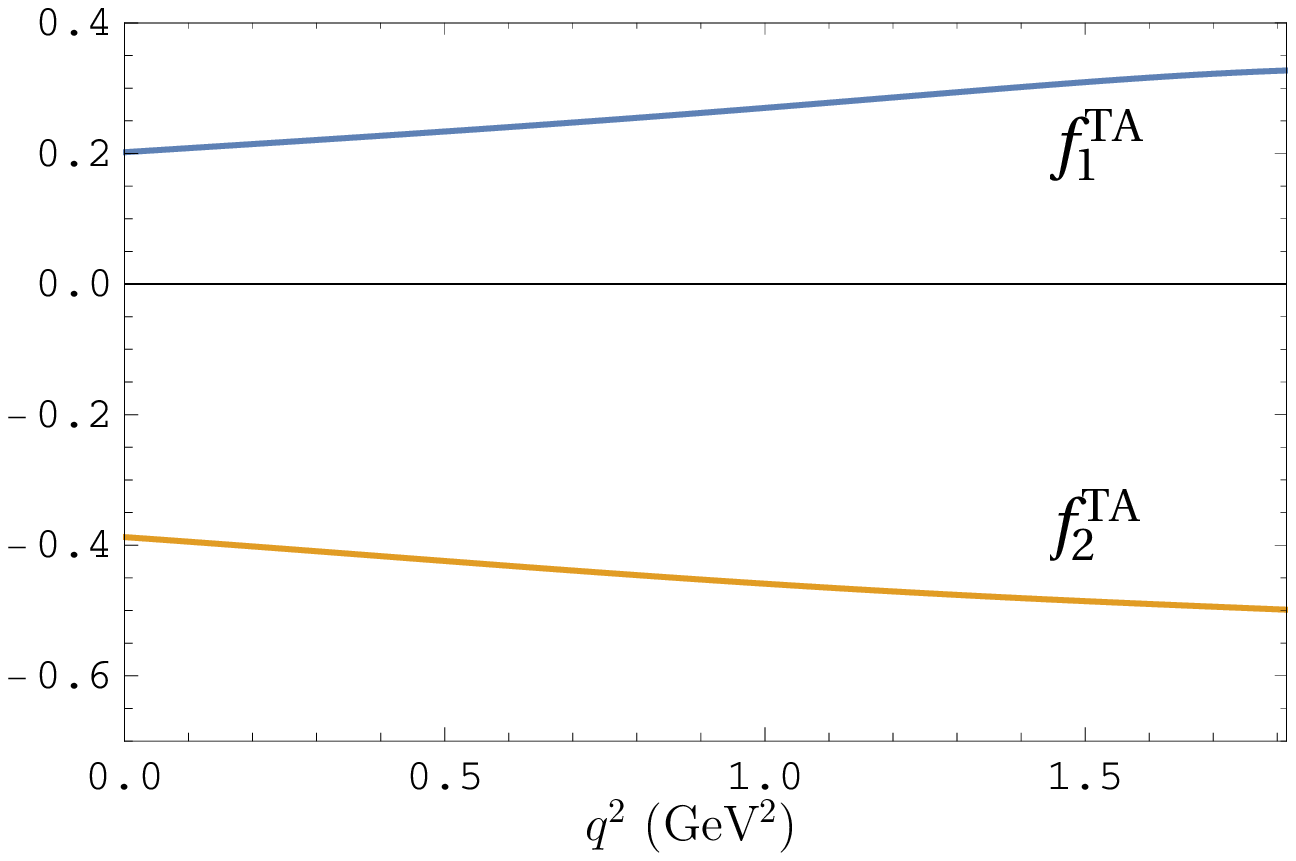}\\
\caption{Form factors of the rare $\Lambda_c\to p$ transition.    } 
\label{fig:ffLcp}
\end{figure}

The calculated form factors are well approximated by the following expressions 
\begin{equation}
  \label{fitff}
  F(q^2)= \frac{1}{{1-q^2/{M_{\rm pole}^2}}} \left\{ a_0 + a_1 z(q^2) +
    a_2 [z(q^2)]^2 \right\},
\end{equation}
where the variable 
\begin{equation}
z(q^2) = \frac{\sqrt{t_+-q^2}-\sqrt{t_+-t_0}}{\sqrt{t_+-q^2}+\sqrt{t_+-t_0}},
\end{equation}
$t_+=(M_D+M_\pi)^2$ and $t_0 = q^2_{\rm max} = (M_{\Lambda_c} - M_{p})^2$.  The pole
masses have the  values: $M_{\rm
  pole}\equiv M_{D^*}=2.010$ GeV for $f_{1,2}^V$, $f_{1,2}^{TV}$; $M_{\rm
  pole}\equiv M_{D_{1}}=2.423$ GeV for $f_{1,2}^A$, $f_{1,2}^{TA}$; $M_{\rm
  pole}\equiv M_{D_{0}}=2.351$ GeV for $f_{3}^V$;  $M_{\rm
  pole}\equiv M_{D}=1.870$ GeV for $f_{3}^A$.
The fitted values of the parameters $a_0$, $a_1$, $a_2$ as well as the
values of form factors at maximum $q^2=0$ and zero recoil $q^2=q^2_{\rm
  max}$ are given in Table~\ref{ffLcp}. The difference of the approximated
form factors from the calculated ones does not exceed 0.5\%. Our model form factors
are plotted in Fig.~\ref{fig:ffLcp}.

\begin{table}
\caption{Form factors of the weak $\Lambda_c\to p$ transition. }
\label{ffLcp}
\begin{ruledtabular}
\begin{tabular}{ccccccccccc}
& $f^V_1(q^2)$ & $f^V_2(q^2)$& $f^V_3(q^2)$& $f^A_1(q^2)$ & $f^A_2(q^2)$ &$f^A_3(q^2)$& $f^{TV}_1(q^2)$ & $f^{TV}_2(q^2)$& $f^{TA}_1(q^2)$ & $f^{TA}_2(q^2)$\\
\hline
$f(0)$          &0.627 &$0.259$ & $0.179$ & 0.433 & $-0.118$&$-0.744$&$-0.310$ & $-0.380$ & 0.202 & $-0.388$\\
$f(q^2_{\rm max})$&0.821  &$0.640$ & $0.303$ & 0.517& $-0.443$& $-1.63$ &$-0.517$ & $-0.505$ & 0.327& $-0.499$\\
$a_0$      &$0.451$&$0.348$& $0.209$& $0.349$ &$-0.309$&  $-0.768$&$-0.285$& $-0.277$& $0.228$ &$-0.342$\\
$a_1$      &$1.51$&$-0.344$&$0.245$&$0.503$&$1.04$&$-0.446$&$-0.915$&$-0.757$&$0.097$&$-0.677$\\
$a_2$      &$-2.12$&$-1.64$& $-3.06$& $0.614$ &$1.72$&  $3.59$&$5.05$& $0.395$& $-1.87$ &$2.49$\\
\end{tabular}
\end{ruledtabular}
\end{table}

We compare our results for the form factors $f^{V,A,TV,TA}_i$  at the
maximum recoil point $q^2=0$ with the predictions of other approaches
in Table~\ref{compbpiff}. The covariant confined  quark model was used
in Ref.~\cite{gikls}. The authors of Refs.~\cite{lhw,kkmw}
employ the QCD light-cone sum rules. The calculations in
Ref. \cite{abss} are based on  full QCD sum rules at light
cone. We find reasonable agreement with results of
Refs. \cite{gikls,kkmw}, while predictions of
Ref. \cite{abss} are substantially different for most of form
factors. Note that the tensor  $f^{TV,TA}_{1,2}$ form factors were not
calculated in Refs. \cite{gikls,kkmw,abss}. It is not possible to
present tensor form factors from the light-cone QCD sum rules
\cite{abss1}, since the authors use a different parametrization for
the matrix element involving tensor current which contains two extra
form factors (6 instead of usual 4).

\begin{table}
\caption{Comparison of theoretical predictions for the form factors of 
  weak baryon $\Lambda_c\to p$ decays at maximum
  recoil point $q^2=0$.  }
\label{compbpiff}
\begin{ruledtabular}
\begin{tabular}{ccccccccccc}
&$f^V_1(0)$&$f^V_2(0)$&$f^V_3(0)$&$f^A_1(0)$&$f^A_2(0)$&$f^A_3(0)$& $f^{TV}_1(0)$ & $f^{TV}_2(0)$& $f^{TA}_1(0)$ & $f^{TA}_2(0)$\\
\hline
our& 0.627& 0.259& 0.179& 0.433& $-0.118$& $-0.744$&$-0.310$ & $-0.380$ & 0.202 & $-0.388$\\
\cite{gikls}&0.470&0.246& $0.039$& 0.414& $-0.073$&$-0.328$\\
\cite{kkmw}&$0.59^{+0.15}_{-0.16}$&$0.43^{+0.13}_{-0.12}$&
&$0.55^{+0.14}_{-0.15}$ &$-0.16^{+0.08}_{-0.05}$&\\
\cite{abss}&0.17&1.78&2.95&0.52&0.71&$-0.0073$\\
\end{tabular}
\end{ruledtabular}
\end{table}

\begin{figure}
\centering
  \includegraphics[width=8cm]{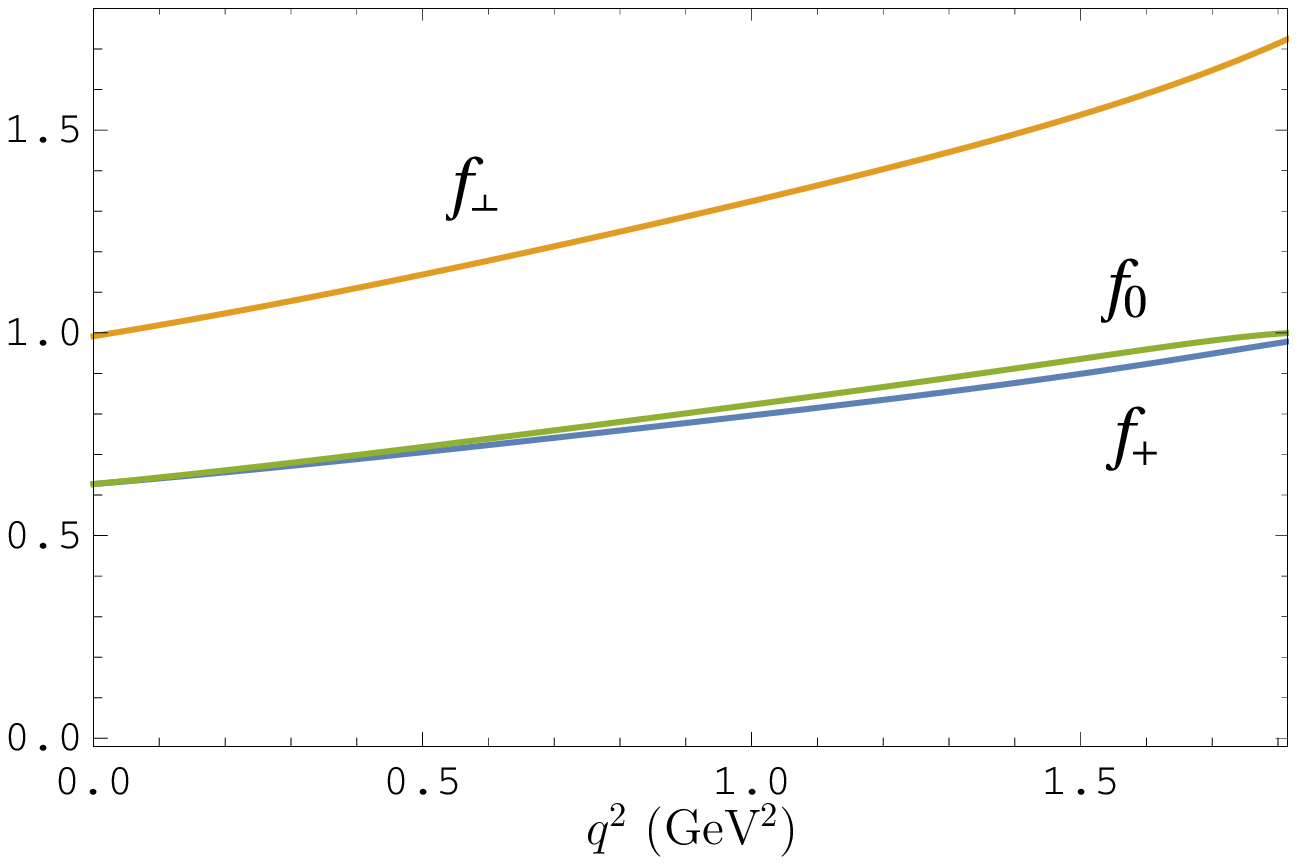}\ \
 \ \includegraphics[width=8cm]{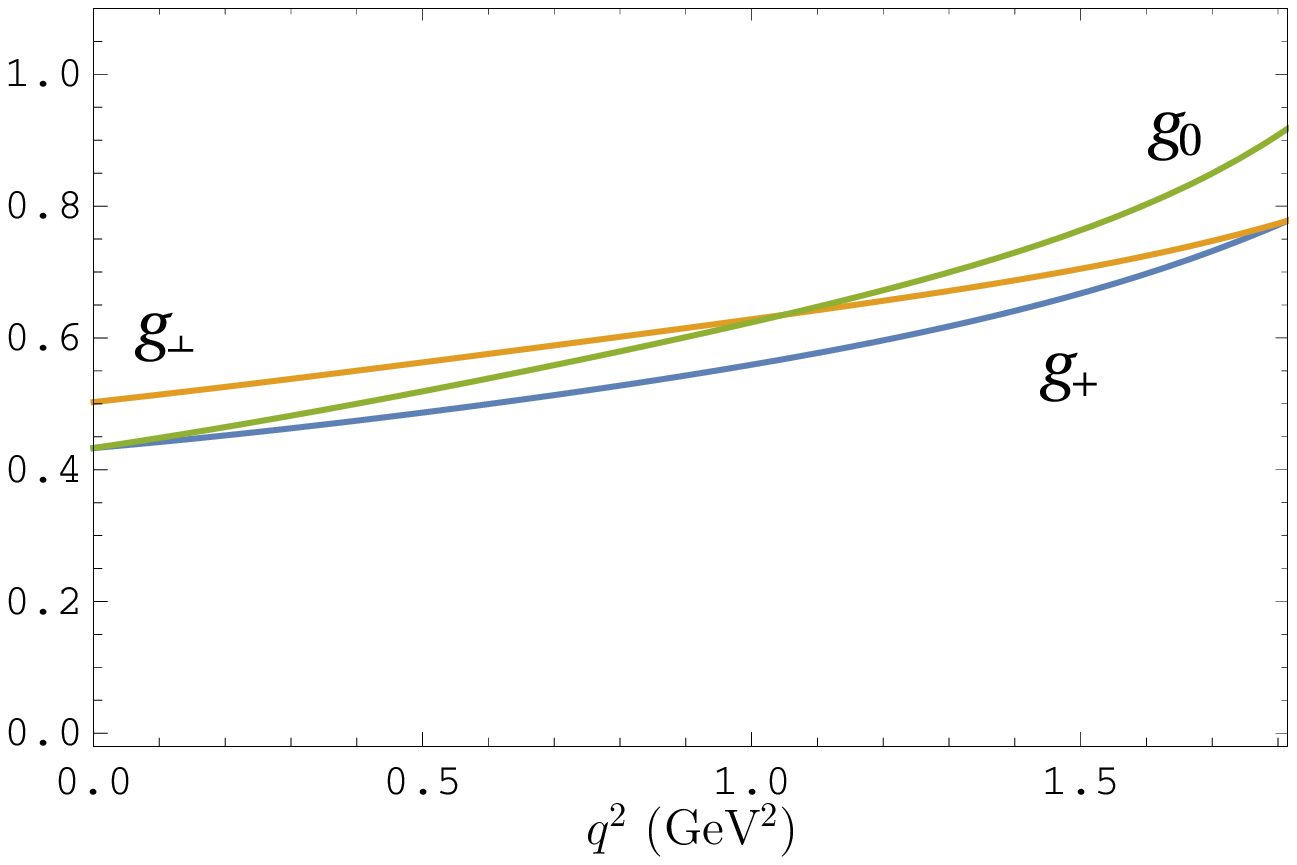}\\
\includegraphics[width=8cm]{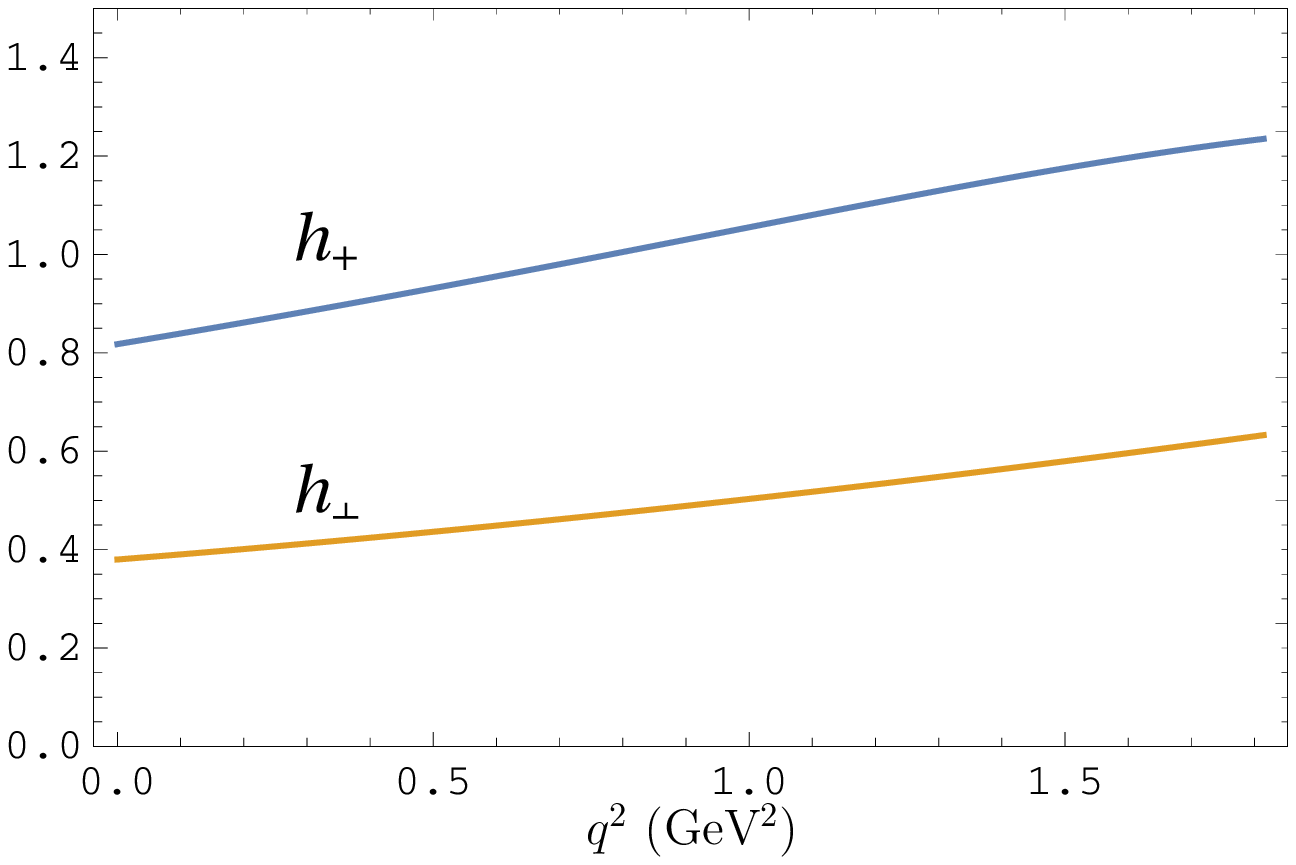}\ \
 \ \includegraphics[width=8cm]{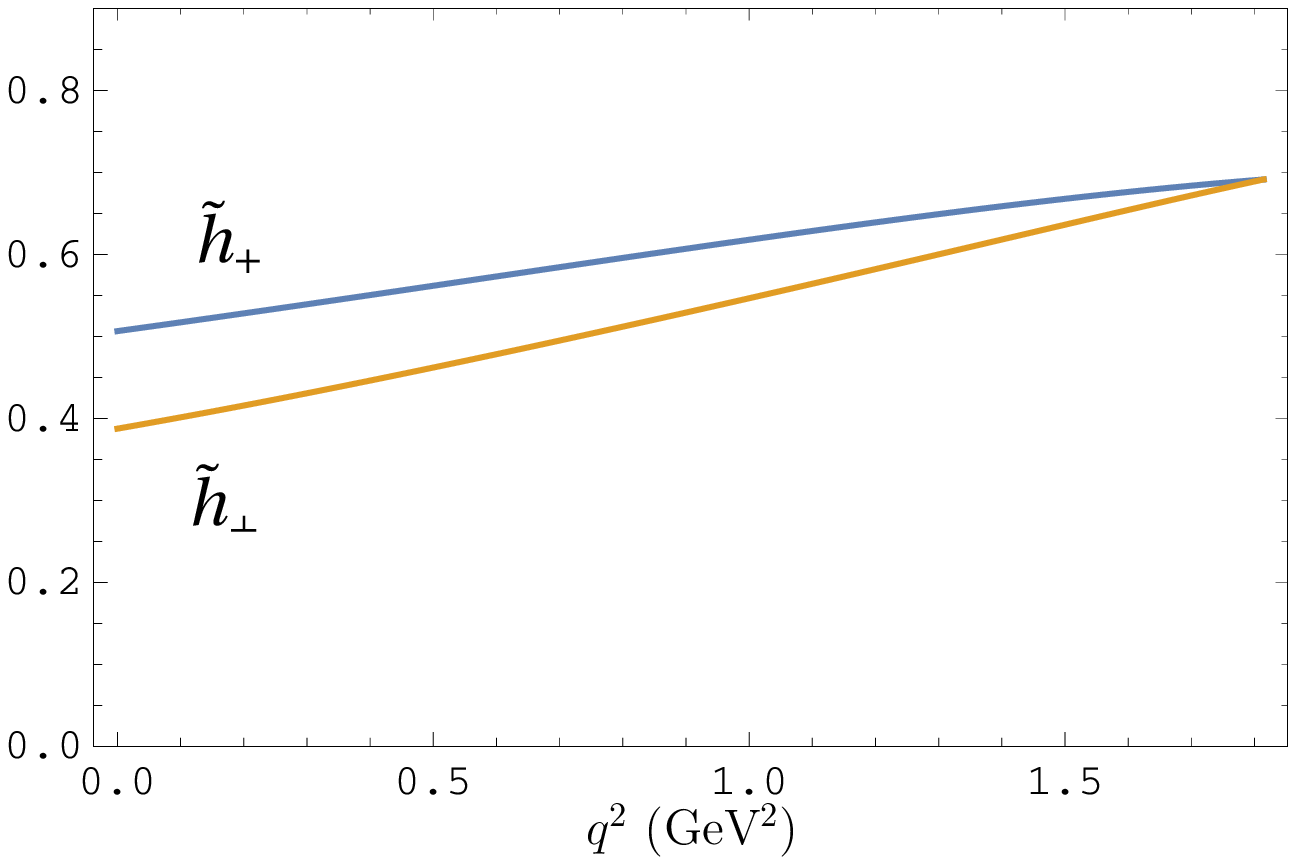}\\
\caption{Helicity form factors of the rare $\Lambda_c\to p$ transition.    } 
\label{fig:hffLcp}
\end{figure}

\begin{table}
\caption{Comparison of our predictions with lattice data for the
  helicity form factors of  weak baryon $\Lambda_c\to p$ decays at
  minimum $q^2=q^2_{\rm max}$ and maximum $q^2=0$
  recoil.  }
\label{compbpqmax}
\begin{ruledtabular}
\begin{tabular}{ccccccccccc}
&$f_+(q^2)$&$f_\perp(q^2)$&$f_0(q^2)$
&$g_+(q^2)$&$g_\perp(q^2)$&$g_0(q^2)$&                    $h_+(q^2)$
  & $h_\perp(q^2)$& $\tilde h_+(q^2)$ & $\tilde h_\perp(q^2)$\\
\hline
$q^2=q^2_{\rm max}$\\
our& 0.978& 1.72& 0.999& 0.778& $0.778$& $0.918$&$1.23$ & $0.633$ & 0.691 & $0.691$\\
\cite{latt}&$1.50(8)$&$2.47(12)$& $1.25(6)$& $0.99(3)$&
                                                        $0.99(3)$&$1.52(8)$&$2.01(14)$&$1.14(6)$& $0.91(4)$& $0.91(4)$\\
$q^2=0$\\
our& 0.627& 0.992& 0.627& 0.433& $0.503$& $0.433$&$0.818$ & $0.380$ & 0.507 & $0.388$\\
\cite{latt}&$0.67(6)$&$1.12(12)$& $0.67(6)$& $0.60(6)$& $0.60(5)$&$0.60(6)$&$0.94(9)$&$0.50(5)$& $0.54(6)$& $0.52(6)$\\
\end{tabular}
\end{ruledtabular}
\end{table}

Recently the form factors of the weak $\Lambda_c\to p$ transitions
were calculated on the lattice \cite{latt}. The author used the
helicity-based definition of the form factors. The relation between
these form factors and the ones defined by Eqs. (\ref{eq:ff}) are
given in Ref.~\cite{lbrare}. We plot the  helicity form factors of the
rare $\Lambda_c\to p$ transition obtained in our model in
Fig.~\ref{fig:hffLcp}. In Table~\ref{compbpqmax} we compare our
results for these form factors with the values calculated on the
lattice \cite{latt} both  at minimum $q^2=q^2_{\rm max}$ and maximum $q^2=0$
recoil of the final proton. We find that our and lattice form factors
at  $q^2=0$ have close values, while most of the lattice form factors
at $q^2=q^2_{\rm max}$ have somewhat larger values than ours. Although
we find the similar behaviour of the form factors, the lattice ones
in general grow more rapidly than ours with the growth of the $q^2$.

\section{Rare $\Lambda_c$ decays}

The low-energy effective Hamiltonian for the $c\to u$ transitions can be
written as follows \cite{fms}
\begin{equation}
  \label{eq:cuheff}
    {\cal H}_{\rm eff} =-\frac{4G_F}{\sqrt{2}}\left(V_{cb}^*V_{ub}{\cal H}_{\rm eff}^{(b)}+V_{cd}^*V_{ud}{\cal H}_{\rm eff}^{(d)}\right),
\end{equation}
where 
\begin{eqnarray*}
  {\cal H}_{\rm eff}^{(b)}&=&c_1{\cal O}_1^s+c_2{\cal O}_2^s+\sum_{i=3}^{10}c_i{\cal
      O}_i,\cr
 {\cal H}_{\rm eff}^{(d)}&=&c_1({\cal O}_1^s-{\cal O}_1^d)+c_2({\cal O}_2^s-{\cal O}_2^d),
\end{eqnarray*}
and  $G_F$ is the Fermi constant, $V_{ij}$ denote the
Cabibbo-Kobayashi-Maskawa matrix elements, $c_i$ are the Wilson coefficients
and ${\cal O}_i^{(q)}$ are the standard model  operators which
expressions can be found e.g. in Ref.~\cite{fms}.

Then the matrix element of the $c\to u l^+l^-$ transition amplitude
between baryon states can be written in the form similar to the $b\to
s l^+l^-$ transition \cite{lbrare}
\begin{equation}
  \label{eq:mtl}
  {\cal M}(\Lambda_c\to p l^+l^-)=\frac{G_F\alpha}{2\sqrt{2}\pi}
 \left[T^{(1)}_\mu(\bar l\gamma^\mu l)+T^{(2)}_\mu
    (\bar l\gamma^\mu \gamma_5l)\right],
\end{equation}
where 
\begin{eqnarray}
  \label{eq:amp}
  T^{(1)}_\mu&=&c_9^{\rm eff}(q^2)\langle p|\bar u
                 \gamma^\mu(1-\gamma^5)c|\Lambda_c\rangle
                 -\frac{2m_c}{q^2}c_7^{\rm eff}(q^2)\langle p|\bar u
                 i\sigma^{\mu\nu}q_\nu(1+\gamma^5)c|\Lambda_c\rangle,\cr
 T^{(2)}_\mu&=&V_{cb}^*V_{ub}c_{10}\langle p|\bar u
                 \gamma^\mu(1-\gamma^5)c|\Lambda_c\rangle.
\end{eqnarray}
The expressions for $T^{(m)}$ ($m=1,2$) in terms of the form factors $f^{V,A,TV,TA}_i$
and the Wilson coefficients are given in Ref.~\cite{lbrare}. 

The effective Wilson coefficient $c_9^{\rm eff}$ contains additional
effects from quark loops which can be treated within perturbation theory as well
as long-distance contributions of the vector resonances $\rho,\omega$ and $\phi$
decaying in to lepton pair $l^+l^-$. The perturbative part is given by
\cite{bms}
\begin{eqnarray}
  \label{eq:c9eff}
  c_9^{\rm
  eff}(q^2)&=&\left({V_{cd}^*V_{ud}}+{V_{cs}^*V_{us}}\right)
             \Biggl[c_9 +h(m_c, q^2)\left(7c_3 + \frac43c_4 + 76c_5 +
             \frac{64}3 c_6\right)\cr
&&-h(m_s, q^2) (3 c_3 + 30 c_5)+\frac43 h(0,q^2) \left(3c_3 + c_4 +
   \frac{69}2 c_5 + 16 c_6\right) + \frac83 (c_3 + 10 c_5)\Biggr]\cr
&&-\left(V_{cd}^*V_{ud} h(0,q^2)+V_{cs}^*V_{us}h(m_s, q^2)\right)\left(\frac23 c_1 + \frac12 c_2\right),
\end{eqnarray}
where
\begin{eqnarray*} 
h(m_q,q^2) & = & 
- \frac{8}{9}\ln\frac{m_q}{m_c} +
\frac{8}{27} + \frac{4}{9} x 
-  \frac{2}{9} (2+x) |1-x|^{1/2} \left\{
\begin{array}{ll}
 \ln\left| \frac{\sqrt{1-x} + 1}{\sqrt{1-x} - 1}\right| - i\pi, &
 x \equiv \frac{4 m_q^2}{ q^2} < 1,  \\
 & \\
2 \arctan \frac{1}{\sqrt{x-1}}, & x \equiv \frac
{4 m_q^2}{q^2} > 1,
\end{array}
\right. \\
h(0, q^2) & = & \frac{8}{27} - 
\frac{4}{9} \ln\frac{q^2}{m_c^2} + \frac{4}{9} i\pi.
\end{eqnarray*}
From these expressions it is clearly seen that the perturbative
contribution is strongly GIM suppressed. In the following calculations
we take the values of the
Wilson coefficients $c_i$ form Ref.~\cite{bms} and the effective Wilson
coefficient $c_7^{\rm eff}(q^2)$ from Ref.~\cite{bh}.

The contributions from the resonances can be modeled by the effective
Wilson coefficient \cite{bh}
\begin{equation}
  \label{eq:cr}
  c_9^{\rm R}(q^2)=a_\rho
  e^{i\delta_\rho}\left(\frac1{q^2-M_\rho^2+iM_\rho\Gamma_\rho}-\frac13\frac1{q^2-M_\omega^2+iM_\omega\Gamma_\omega}\right)+
  \frac{a_\phi e^{i\delta_\phi}}{q^2-M_\phi^2+iM_\phi\Gamma_\phi},
\end{equation}
where $M_M$ and $\Gamma_M$ are masses and total widths of vector
$M=\rho,\omega,\phi$ mesons which we take form Ref.~\cite{pdg}. The
isospin relation between the $\rho$ and $\omega$ contributions was
explicitly taken into account. The coupling $a_\phi$ can
be determined from the experimental data  \cite{pdg}
\begin{eqnarray}\label{eq:brphi}
  Br(\Lambda_c\to p\phi)&=&(1.08\pm0.14)\times 10^{-3},\cr
Br(\phi\to\mu^+\mu^-)&=&(2.87\pm0.19)\times 10^{-4},
\end{eqnarray}
and  $a_\rho$ from the recent LHCb measurement \cite{lhcblc} of the
ratio
\begin{equation}\label{eq:bromega}
\frac{Br(\Lambda_c\to p\omega)Br(\omega\to \mu^+\mu^-)}{Br(\Lambda_c\to p\phi)Br(\phi\to\mu^+\mu^-)}=0.23\pm0.08\pm0.03 
\end{equation}
by  approximating
\begin{equation}
  \label{eq:brf}
  Br(\Lambda_c\to pV\to p\mu^+\mu^-)= Br(\Lambda_c\to pV)Br(V\to
  \mu^+\mu^-), \qquad (V=\phi,\omega).
\end{equation}
As a result the following values are obtained
\begin{eqnarray*}
  a_\rho&=&(0.21\pm0.04)\ {\rm GeV^2},\cr
a_\phi&=&(0.13\pm0.01)\ {\rm GeV^2}.
\end{eqnarray*}
These coefficients have close values to the ones found previously within
the analysis of the rare $D$ meson decays \cite{bh}. The relative
strong phases $\delta_\rho$ and $\delta_\phi$ are not known and we
vary them independently in 0 to $2\pi$ range.

The lepton angle differential decay distribution is given by
\begin{equation}
  \label{eq:ddGl}
  \frac{d^2 \Gamma(\Lambda_c\to p l^+l^-)}{d q^2d\cos\theta}=\frac{d
    \Gamma(\Lambda_c\to p l^+l^-)}{d q^2}\left[\frac38(1+\cos^2\theta)(1-F_L)+ A_{FB}^\ell\cos\theta+\frac34F_L\sin^2\theta\right],\qquad
\end{equation}
where $\theta$ is the angle between the $\Lambda_c$ baryon and the positively
charged lepton in the dilepton rest frame, $A_{FB}^\ell$ is the lepton
forward-backward asymmetry and $F_L$ is the fraction of the
longitudinally polarized dileptons. The explicit expressions for the
differential decay rates, forward-backward asymmetry and  the fraction of the
longitudinally polarized dileptons in terms of form factors can be
straightforwardly obtained from the ones given in Ref.~\cite{lbrare}. 

Substituting in these expressions the
decay form factors calculated within our model in the previous section we
get predictions for the rare $\Lambda_c\to p l^+l^-$ decay
observables. We calculate them separately both without and with the inclusion of
the long-distance vector $\rho,\omega,\phi$ resonance contributions
(LD). The obtained results for the branching fractions are given in
Table~\ref{brlc} in comparison with other theoretical predictions. We
see that our predictions agree well with recent lattice QCD
calculations \cite{latt} especially for the  $\Lambda_c\to p\mu^+\mu^-$ branching
fraction with the account of the resonance contributions. On the other
hand,  the nonresonant branching fractions from Refs.~\cite{s,abss1},
which employ the light-cone sum rules, are  2 orders of
magnitude lower than our predictions and 3 orders of magnitude lower
than the lattice results \cite{latt}. 
Our prediction for the $\Lambda_c\to p\mu^+\mu^-$ nonresonant
branching fraction is about a factor of 15 lower than the lattice QCD
result  \cite{latt}. Such deviation cannot be explained entirely by
the above mentioned differences in the form factors, and could be additionally caused by the
different choices for the perturbative effective Wilson
coefficients. Thus lattice calculation includes the two-loop matrix
elements of $O_1$ and $O_2$ from Ref.~\cite{sdb} which we do not take
into account in the present consideration.    
 We roughly estimate
the uncertainty of our predictions for the nonresonant differential
and total branching fractions to be about 15\%. Its sources are the
following: 10\% comes from the uncertainties of our form factors and additional
5\% emerges from the variation of the renormalization scale $\mu$ of the Wilson
coefficients from $\mu=1.3$~GeV \cite{bms} to $\mu=1.5$~GeV
\cite{fms}.\footnote{In fact, this uncertainty could be significantly
  larger if we allow the wider range of the renormalization scale
  $\mu$ variation, e.g., $m_c/\sqrt2\le \mu\le \sqrt2 m_c$ (see
  discussion in Ref.~\cite{bh}).}  
Note that our value of the
$\Lambda_c\to p\mu^+\mu^-$ nonresonant branching fraction is very
close to the recent prediction \cite{bh} for the corresponding nonresonant
branching fraction of the $D^+\to \pi^+\mu^+\mu^-$ decay.

\begin{table}
\caption{Comparison of theoretical predictions for the $\Lambda_c$ rare decay
  branching fractions. }
\label{brlc}
\begin{ruledtabular}
\begin{tabular}{ccc}
& nonresonant & with resonances (LD) \\
\hline
$\Lambda_c\to p e^+e^-$\\
our& $(3.8\pm0.5)\times 10^{-12}$&  $(3.7\pm0.8)\times 10^{-7}$\\
\cite{s}&  $(4.5\pm2.37)\times 10^{-14}$& $(4.2\pm0.73)\times
                                          10^{-6}$\\
\cite{abss1}&  $(4.19\pm2.35)\times 10^{-14}$&\\
$\Lambda_c\to p \mu^+\mu^-$\\
our& $(2.8\pm0.4)\times 10^{-12}$&  $(3.7\pm0.8)\times 10^{-7}$\\
\cite{latt}& $(4.1\pm0.4^{+6.1}_{-1.9})\times 10^{-11}$&  $(3.7\pm1.3)\times 10^{-7}$\\
\cite{s}&  $(3.77\pm2.28)\times 10^{-14}$& $(3.2\pm0.66)\times
                                          10^{-6}$\\
\cite{abss1}&  $(3.87\pm2.26)\times 10^{-14}$&\\
\end{tabular}
\end{ruledtabular}
\end{table}      

\begin{figure}
  \centering
 \includegraphics[width=12cm]{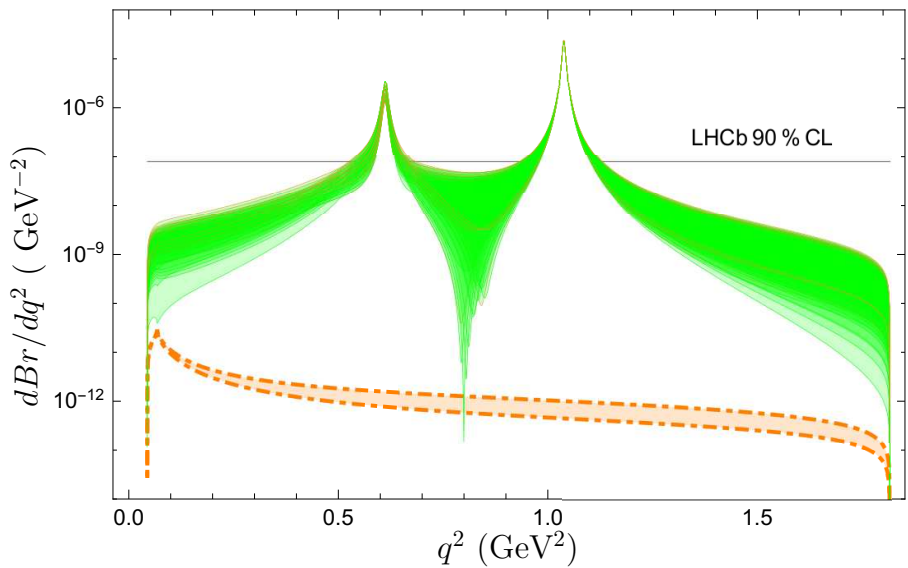}
  \caption{The differential branching fractions for the
    $\Lambda_c\to p\mu^+\mu^-$ rare decay. The area between
    dashed-dotted (orange) curves corresponds to the nonresonant predictions,
    while the (green) band shows resonance contributions including
    uncertainties in  coefficients $a_\rho$ and $a_\phi$ as well as
    variations of the relative strong phases $\delta_\rho$ and
    $\delta_\phi$. The horizontal black line denotes the LHCb 90\% CL upper limit \cite{lhcblc}.}
  \label{fig:brLc}
\end{figure}

Our result for the branching fraction of the  $\Lambda_c\to p e^+e^-$
rare decay is well below the experimental upper limit set by the
BABAR Collaboration $Br(\Lambda_c^+\to p e^+e^-)<5.5\times 10^{-6}$ \cite{babarLc}.

In Fig.~\ref{fig:brLc} we plot our predictions for the differential
branching fractions of the $\Lambda_c\to p\mu^+\mu^-$ rare decay both
without and with inclusion of the vector $\rho,\omega,\phi$ meson
contributions. The LHCb 90\% CL upper limit \cite{lhcblc} is also
given. We see that the calculated branching fraction with the account
of resonances agree well with the experimental limit. If we calculate
the branching fraction in a region with excluded ranges $\pm40$ MeV
around the $\omega$ and $\phi$ masses, we get the value
$(1.9\pm0.5)\times 10^{-8}$, which is in accord with experimental
upper limit $Br(\Lambda_c^+\to p\mu^+\mu^-)<7.7\times 10^{-8}$
obtained with the same constraints. Note that the prediction with the
account of resonances of  Ref.~\cite{s} is significantly above this experimental limit.       

\begin{figure}
  \centering
 \includegraphics[width=12cm]{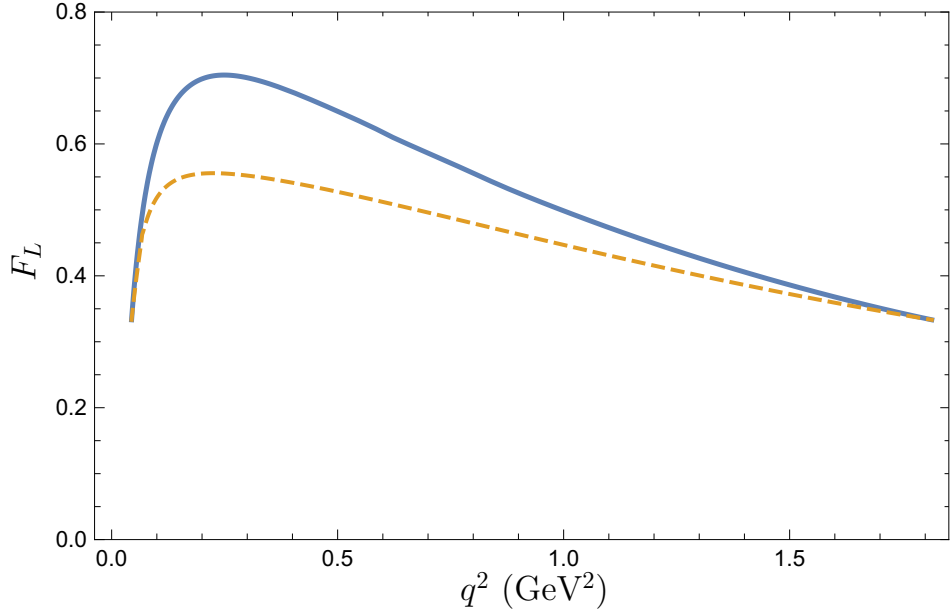}

  \caption{Prediction for the fraction of longitudinally polarized
    dileptons $F_L(q^2)$  in the  $\Lambda_c\to p \mu^+\mu^-$ decay. The  dashed curve corresponds to the nonresonant predictions,
    while the solid curve shows results with the inclusion of the vector meson resonances. }
  \label{fl}
\end{figure}

In Fig.~\ref{fl} we plot our prediction for the $q^2$-dependence of
the fraction of the longitudinally polarized dimuons $F_L(q^2)$ for the
central values of the decay parameters. The predicted value of the
average longitudinal polarization fraction with the account of resonance
contributions is $\left< F_L\right> =0.52\pm 0.02$. The lepton
forward-backward asymmetry $A_{FB}^\ell$ is proportional to the Wilson
coefficient $c_{10}$ and thus vanishes in the standard
model. Therefore, any
experimental deviations from zero for $A_{FB}^\ell$ will be a signal
of the new physics.

\section{Conclusions}

The $\Lambda_c\to p l^+l^-$ rare decays were investigated in the
framework of the relativistic quark model.  The quasipotential
approach with the QCD-motivated interquark interaction was employed
for the calculation of the $\Lambda_c\to p$
weak transition from factors. The
relativistic effects including wave function
transformations of the final proton from the rest to moving reference
frame and contributions to decay processes
of the intermediate negative-energy states are comprehensively taken into account. These form factors were expressed
through the overlap integrals of the baryon wave functions and their
$q^2$ dependence was consistently determined in the whole accessible
kinematical range.  No additional model assumptions and extrapolations
were used thus improving the reliability of the
obtained results. Reasonable agreement of the calculated  helicity
form factors at $q^2=0$ with recent lattice results \cite{latt} is
found. However our form factors increase with growing $q^2$ slightly
slowly than the lattice ones.   

These form factors were used to calculate the differential and total branching
fractions and angular distributions of the $\Lambda_c\to p l^+l^-$
rare decays. Both the perturbative and effective
Wilson coefficients, which include the additional
long-distance contributions from the vector $\rho,\omega$ and $\phi$
resonances were used in the analysis. It was found that the
perturbative term is strongly GIM suppressed and the main contribution
comes from resonances modeled by a simple Breit-Wigner
model, which coefficients are determined from available experimental
data. The calculated branching fraction of the  $\Lambda_c\to
p\mu^+\mu^-$ decay is well consistent with the experimental
upper limit $Br(\Lambda_c^+\to p\mu^+\mu^-)<7.7\times 10^{-8}$ at 90\%
confidence level recently reported by the LHCb \cite{lhcblc}.

\acknowledgments
We are grateful to A. Ali, D. Ebert and M. Ivanov for valuable  discussions and support.

\end{document}